# Biopsies prostatiques sous guidage échographique 3 dimensions et temps réel (4D) sur fantôme. Etude comparative versus guidage 2D.

Jean-Alexandre LONG (1), Vincent DAANEN (2), Alexandre MOREAU-GAUDRY (3), Jocelyne TROCCAZ (2), Jean-Jacques RAMBEAUD (1), Jean-Luc DESCOTES (1)

1- service de chirurgie urologique et de la transplantation rénale CHU Grenoble
2- Laboratoire TIMC-IMAG Grenoble (CNRS 5525)
3- Centre d'innovation technologique CHU Grenoble

Adresse pour correspondance :

Jean-Alexandre LONG

**Service de chirurgie urologique et de la transplantation rénale**

**Hopital Michallon CHU Grenoble**

**38043 Grenoble cedex 09**

**E-mail : JALong@chu-grenoble.fr**

**Tel: 06 09 91 70 02**

Fax : 04 76 76 56 11

# Biopsies prostatiques sous guidage échographique 3 dimensions et temps réel (4D) sur fantôme. Etude comparative versus guidage 2D.

Jean-Alexandre Long, Vincent Daanen, Alexandre Moreau-Gaudry, Jocelyne Troccaz,
Jean-Jacques Rambeaud, Jean-Luc Descotes

**Résumé**

**Objectif:** L'objectif de l'étude est de définir la valeur ajoutée du guidage échographique temps réel en 3 dimensions (4D) des biopsies sur un fantôme de prostate en terme de précision de localisation et de distribution.

**Méthodes** : Un fantôme de prostate a été réalisé. Un échographe 3D temps réel couplé à une sonde endorectale volumique 5.9 MHz a été utilisé. Quatorze opérateurs ont réalisé 336 biopsies en 2D puis en 4D selon un protocole 12 biopsies.

Le trajet de biopsies a été modélisé par segmentation dans un volume échographique 3D. Un logiciel spécifique a permis la visualisation des trajets de biopsies dans la prostate de référence et d'évaluer la zone ponctionnée. Une étude comparative a été réalisée afin de déterminer l'intérêt apporté par la ponction en 4 D par rapport au 2D en évaluant la précision des points d'entrée et de cible. La distribution était évaluée par la mesure du volume exploré et par un rapport de redondance des points biopsiés.

**Résultats :** La précision de réalisation selon le protocole était améliorée de façon significative en 4D (p=0,037). Il n'a pas été montré une augmentation du volume biopsié ou une amélioration de la répartition des biopsies en 4D par rapport au 2D.

**Conclusion :** La méthode de biopsies de prostate par guidage échographique 3D temps-réel semble montrer sur modèle synthétique une amélioration dans la précision localisatrice et dans la faculté à reproduire un protocole. La répartition des biopsies ne semble pas améliorée.

# Introduction

L'étude que nous proposons a pour objectif d'évaluer le guidage échographique des biopsies de prostate en 3 dimensions et en temps réel (mode 4D).

Le but est de savoir si la visualisation de plusieurs plans de coupe simultanés permet une ponction plus précise et plus reproductible et si une distribution (répartition) plus homogène des biopsies dans le volume prostatique est effectuée.

Nous réalisons une étude comparative entre la ponction classique 2D et la ponction assistée par l'échographie 3D temps réel (4D) sur un modèle synthétique (fantôme).

De nombreux protocoles de biopsies ont été mis au point afin d'améliorer la sensibilité d'une série de biopsies tout en limitant le nombre de ponctions. La tendance est à l'augmentation du nombre de biopsies et à un échantilonage de la zone périphérique. La morbidité additionnelle des protocoles extensifs est débattue [1].

L'échographie 3D ou volumique est une technique actuellement bien au point et utilisée sur certains appareils depuis 1994. Elle consiste à réaliser non pas un plan de coupe, comme en échographie 2 D classique, mais à reconstruire un volume obtenu grâce à un balayage de plan de coupe jointifs. Le temps de balayage varie de 3 à 10 secondes suivant l'importance du volume et la qualité recherchée. On peut aussi utiliser des sondes endocavitaires, dans ce cas le balayage peut être soit rotatif dans l'axe de la sonde, soit angulaire et c'est le cas de l'appareil testé (General Electric Voluson 730 pro ®) . On peut réaliser une acquistion d' un volume 3D d'un organe permettant de manipuler les images et de garder en mémoire les coordonnées de tous les points du volume avec leur niveau de gris correspondant.

Les dernières évolutions des échographes concernent l'acquisition en 3D temps réel, aussi appelé mode 4D. Il consiste à afficher en temps réel 3 plans de coupe orthogonaux permettant un guidage dans 3 plans simultanés.

Ce mode de guidage a fait la preuve de son efficacité en terme de précision dans les ponctions de tumeurs du sein [2], de tumeurs du foie [3] et de pseudo-kystes pancréatiques [4]. Nous voulons étudier son intérêt dans les biopsies de prostate.

**Méthodes**

L'appareil utilisé est un VOLUSON® 730 Pro (General Electric) équipé d'une sonde volumique endorectale RIC5-9 utilisable en mode 2D,3D et 4D grâce à un balayage de 2 plans de coupe orthogonaux. Le transducteur est placé à l'extrémité de la sonde. La fréquence de 5,9 Mhz permet de réaliser des échographies prostatiques et des biopsies.

Il nous a paru être une étape préliminaire de tester l'appareil en premier lieu sur un modèle synthétique.

### Création d'un modèle synthétique de prostate (fantôme)

Dans une boite plastique carrée de 10 cm de côté, un orifice circulaire de 6 cm de diamètre a été réalisé permettant l'introduction de la sonde. A l'intérieur de cet orifice est mis en place un cylindre simulant la paroi rectale. Le milieu de propagation des ultrasons est réalisé par du gel échographique Aquasonic® (750 ml par fantôme). La prostate de 40 ml est réalisée en coulant du gel à bougie dans un moule de silicone (figure 1).

Après biopsie, le trajet était visible en échographie grâce à l'air introduit par l'aiguille. Ceci a permis une évaluation du trajet de biopsies par une acquisition volumique (3D) échographique. Le même modèle de fantôme a été réalisé pour chaque biopsie (figure 2).

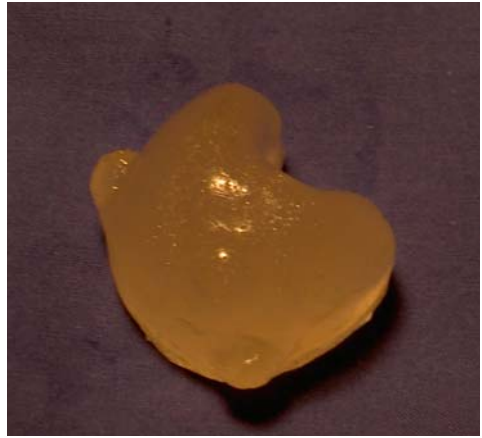

**Figure 1: Le fantôme de prostate en gel de bougie**

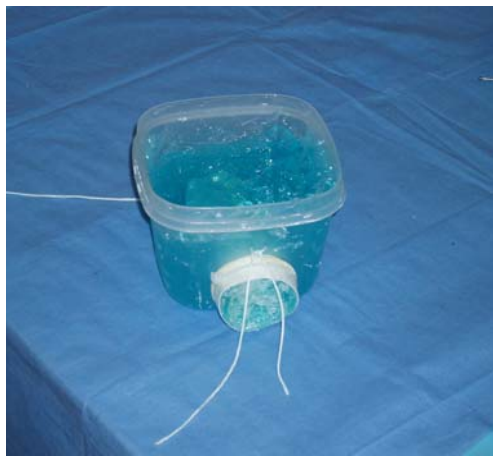

**Figure 2: Le fantôme de biopsie de prostate endorectale**

### Modélisation du trajet de biopsie

L' air incorporé dans le fantôme par le passage de l'aiguille laissait une trace hyperéchogène aisément détectable par les ultrasons (figure 3). On réalisait une acquisition d'un volume 3D de l'ensemble du fantôme au sein duquel se trouvait la biopsie.

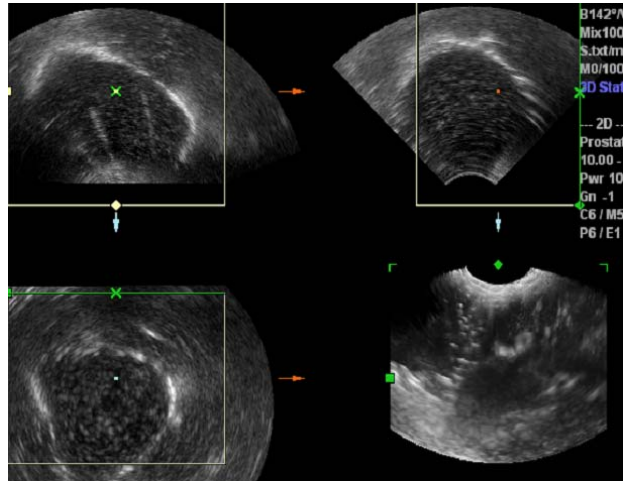

**Figure 3: Visualisation de plusieurs traces de biopsies dans le même fantôme**

**Protocole des biopsies**

Le protocole utilisé a été un protocole 12 biopsies réalisant 2 biopsies (paramédiane et latérale) à l'apex, au milieu et à la base de chaque lobe prostatique.

Compte-tenu du biais qu'auraient représentés les traces des biopsies précédentes, il était indispensable de réaliser un fantôme par biopsie.

Il a donc été réalisé 12 fantômes permettant une séance ininterrompue. Les fantômes étaient remplacés après chaque séance et 336 fantômes ont été réalisés.

Chaque opérateur effectuait une première séance de biopsies avec le VOLUSON 730® en mode 2D comme habituellement puis en mode 4D. Après chaque biopsie, un volume échographique était acquis et mis en mémoire.

### Traitement des données ultrasonographiques acquises.

La première étape du traitement informatique des biopsies a été de créer un volume de référence (modèle ou template) du fantôme.

Chaque fantôme contenant une seule biopsie a fait l'objet d'une acquisition volumique.

Les volumes ont été recalés entre eux à l'aide d'un recalage manuel volumique sur Analyze®.

La recalage ou "fusion" consiste à aligner les volumes en permettant de visualiser les différentes biopsies dans un seul volume.

Des tentatives de recalage automatique ont été infructueuses en raison du bruit important et des variations importantes du contraste des interfaces entre les différents fantômes.

Les douze biopsies d'une séance affichés dans la prostate de référence (figure 4 et 5). La méthodologie est résumée dans la figure 6.

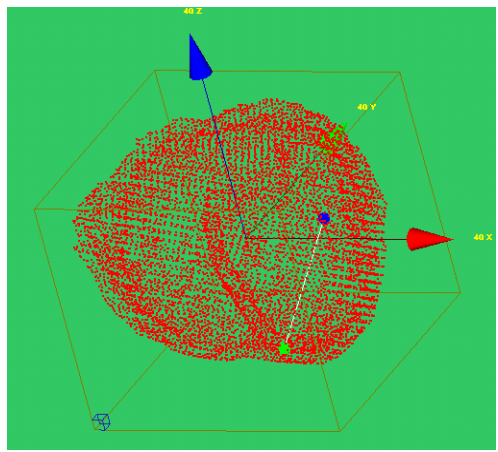

**Figure 4: Visualisation d'un trajet de biopsies dans le volume de référence. La boule verte représente le point d'entrée. La boule bleue représente la cible atteinte.**

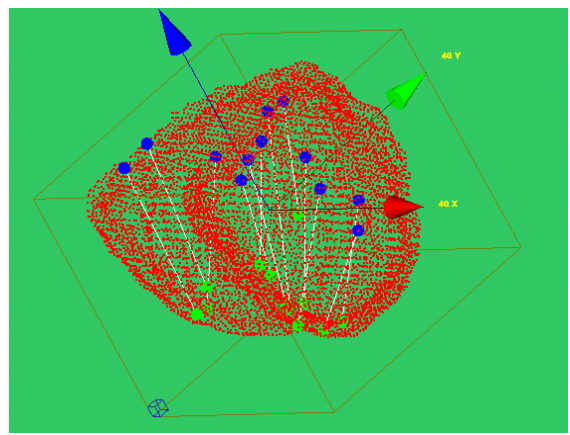

**Figure 5: Superposition de toutes les biopsies dans le template.**

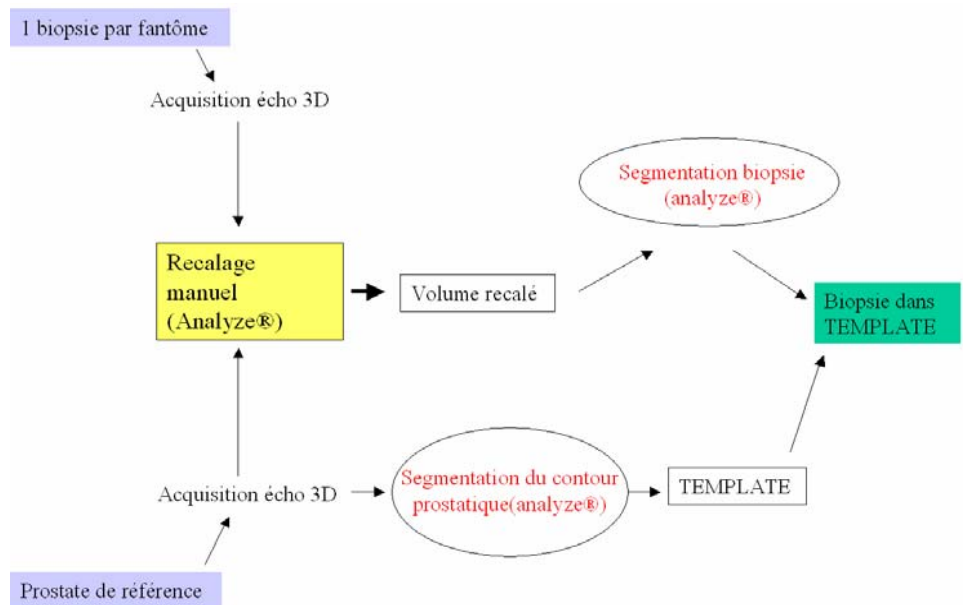

**Figure 6: Schéma récapitulatif de la méthodologie**

*Evaluation des résultats des biopsies*

Un protocole d'évaluation de bonne qualité des biopsies a été établi.Trois paramètres ont été étudiés:

- **Zone d'entrée**: D'après les schémas anatomiques et le schéma du protocole 12 biopsies, un point situé sur la surface prostatique a été choisi comme étant le point moyen idéal d'entrée de l'aiguille dans la prostate pour chaque biopsie. Compte-tenu de l'absence de recommandation claire et de l'absence de cible évidente, une zone d'entrée de 7 mm de diamètre autour de ce point de ponction était considérée comme correcte. Au delà de cette zone, une distance à partir de cette sphère de 7 mm était mesurée. Celle-ci représentait la distance d'erreur. Ce diamètre de 7 mm a été choisi car il représentait une distance permettant une juxtaposition des zones d'entrée sans superposition.

-**Zone cible**: L'extrémité de la ponction était également comparée à une zone cible idéale moyenne de 7 mm de diamètre. L'erreur correspondait à la distance séparant l'extrémité de la biopsie et la sphère (figure 7).

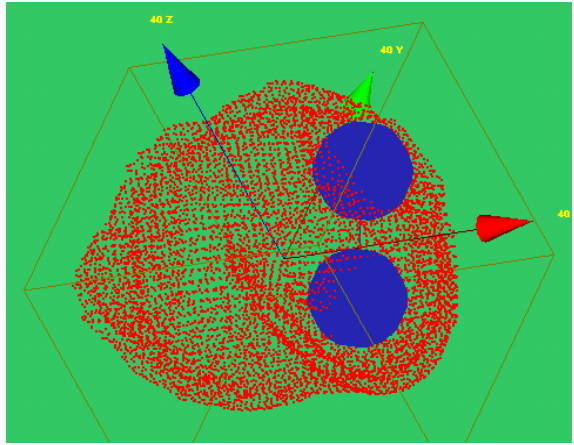

**Figure 7: Création de sphères idéales d'entrée et de cible.**

- **Distribution des biopsies dans le volume**.

Cette évaluation d'une bonne répartition des biopsies dans le volume prostatique n'a pas été simple. Pour Epstein, un cancer est significatif au dessus de 0,5 ml [5]. Nous avons donc essayé de trouver une méthode permettant de mettre en évidence le volume exploré par les biopsie (dépendant de la longueur de biopsie dans la prostate et surtout du recouvrement des biopsies entre elles). L'hypothèse théorique était que 2 biopsies parallèles et distantes de 10mm ne pourraient pas laisser passer un cancer significatif (le diamètre d'une boule de 0,5 cc est 10 mm). Nous avons donc émis l'hypothèse qu'une biopsie explorait un volume compris dans un cylindre de 1,2 mm de diamètre (diamètre de l'aiguille de 18 gauge) auquel s'ajoutait une "zone de sécurité" de 10 mm de diamètre.

La figure 8 montre le suréchantillonage d'une zone en faisant recouvrir 2 volumes cylindriques en rapprochant trop 2 biopsies. Il s'en suit que les régions sont sous-échantillonnées.

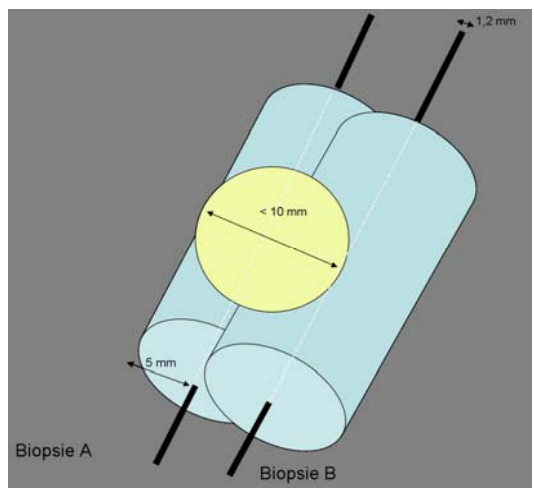

**Figure 8: Chevauchement de 2 "cylindres biopsiques"**

Le volume prostatique exploré par biopsie correspondait donc à la longueur de la biopsie multipliée par la surface d'un disque de rayon 5,6mm.

Les points qui se recoupaient avec les cylindres biopsiques adjacents n'étaient pris en compte qu'une fois.

Nous avons opté pour un ratio: nombre de points dans la prostate explorés/nombre de points qui auraient dû être biopsiés si les biopsies ne se croisaient pas. Cette méthode était un reflet indirect de la distribution en recherchant chez chaque opérateur sa capacité à répartir ces ponctions afin qu'elles ne se recoupent pas.

Au sein de chaque séance, il était possible de déterminer au sein de la prostate le volume recensé par la série.

L'analyse statistique à l'aide du logiciel STATVIEW 5.0®. Les variables quantitatives ont été décrites par leur moyenne et écart-type. Pour la comparaison des moyennes en série appariée, le test statistique de Student pour série appariée a été utilisé, après vérification des conditions d'application.Lorsque ces dernières n'étaient pas vérifiées, un test non paramétrique a été utilisé. Le risque alpha de première espèce aété choisi à 0,05 comme usuellement

Une analyse multivariée en ANOVA concernant les facteurs de localisation dans le plan sagittal (apex, milieu, base) et transversale (paramédian, latéral) a été réalisée.

Le protocole de biopsies standard était expliqué à chaque opérateur par un schéma récapitulatif. La modalité 2D était réalisée en premier. La deuxième séance était effectuée dans un délai de quelques heures à 3 semaines.

## Résultats

Au total, 336 biopsies ont été réalisées par 14 opérateurs. Pour l'analyse des séries appariées, 294 biopsies ont été prises en compte.

Un gain de précision concernant l'atteinte de la zone cible a été retrouvé significativement en 4D par rapport au 2D (p=0,0042) (tableau 1 et 2). Cet apport significatif a été retrouvé dans toutes les localisations de la prostate (tableau 3)).

Par contre on n'a pas retrouvé de gain significatif de précision de ponction de la zone d'entrée (tableau 1 et 2) et ce quelque soit la localisation dans la prostate (tableau 4).

**Tableau 1 : Moyenne des distances aux zones d'entrée et cibles du protocole idéal en modalité 2D et 4D**

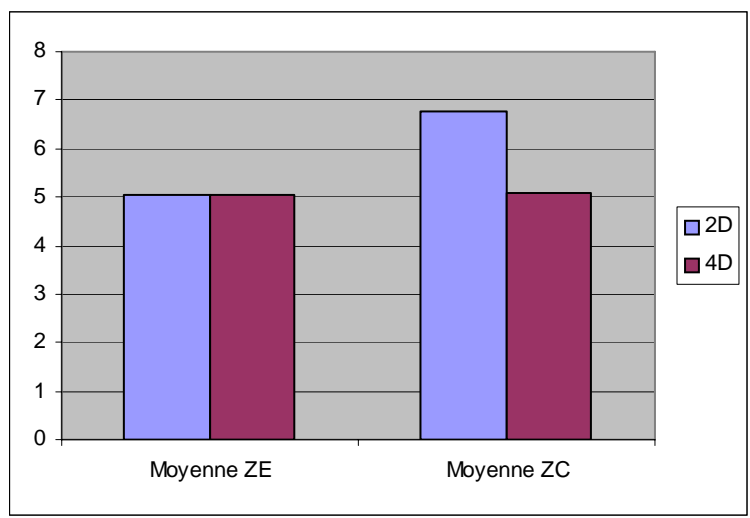

**Tableau 2 : Tableau récapitulatif concernant la précision des biopsies à atteindre le protocole théorique idéal**

|  | n | Modalité | Moyenne (mm) | Ecart-type | Test Student p |
|---|---|---|---|---|---|
| **Distance par rapport à la zone cible idéale** | 147 | **2D** | **6.79** | **6.18** | **0.0042** |
|  |  | **4D** | **5.1** | **4.8** |  |
| **Distance par rapport à la zone d'entrée idéale** | 147 | **2D** | **5.28** | **5.71** | 0.86 |
|  |  | **4D** | **5.19** | **4.83** |  |

**Tableau 3: Courbe des interactions des différences des points d'atteinte de la zone cible en 2D et 4D en fonction des localisations dans la prostate.**

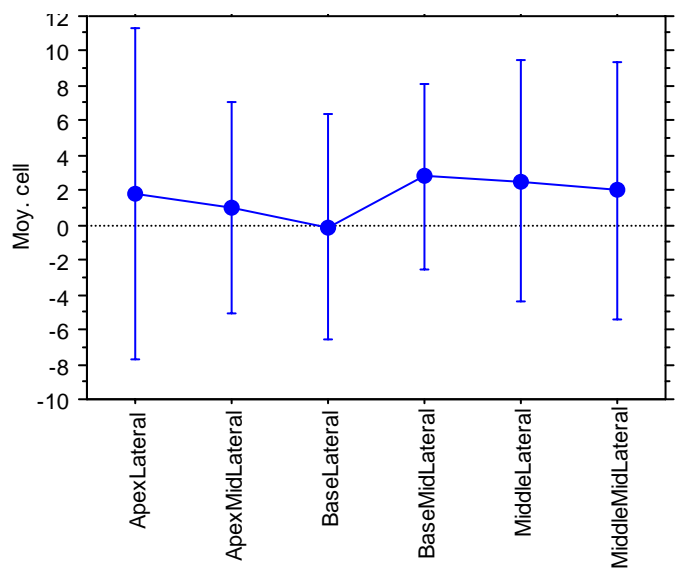

**Tableau 5: Courbe des interactions des différences des points d'entrée en 2D et 4D en fonction des zones de la prostate**

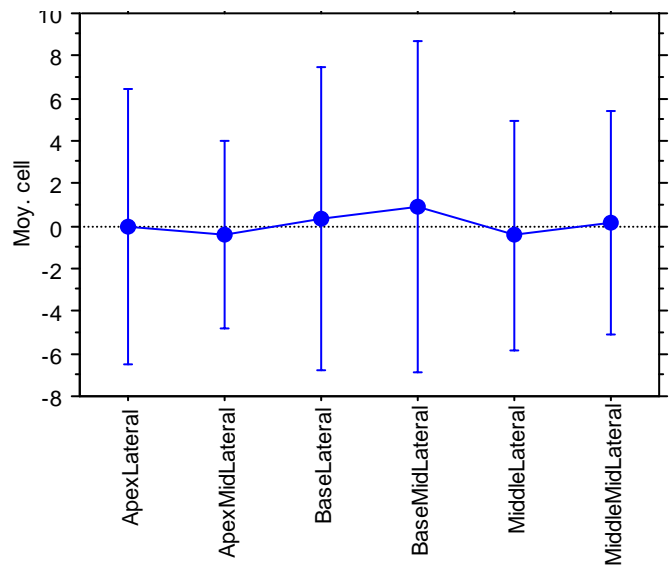

Une étude de reproductibilité des biopsies par l'étude de la variance sur les 294 biopsies des distances de chaque biopsie aux zones d'entrée et de sortie fait resortir une reproductibilité augmentée de manière significative des biopsies vers la zone cible en mode 4D (p= 0,0385). Par contre, il n'existe pas de lien statistique mais une tendance à une amélioration de la reproductibilité des points d'entrée des biopsies dans la prostate (p=0.0623) (tableau 7).

**Tableau 7: Etude de la reproductibilité des biopsies par l'écart à la moyenne² (variance)**

|  | n | Modalité | Ecart à la moyenne (mm) | Ecart-type | Test Student p |
|---|---|---|---|---|---|
| **Variance des distance par rapport à la zone cible idéale** | 147 | **2D** | **37.95** | 78.45 | **0.0385** |
|  |  | **4D** | **22.9** | 41.38 |  |
| **Variance des distance par rapport à la zone d'entrée idéale** | 147 | **2D** | **32.43** | 59.66 | 0.0623 |
|  |  | **4D** | **23.18** | 33.15 |  |

L'étude de la distribution des biopsies n'a pas retrouvé de différence significative de volume exploré dans la prostate en 2D et en 4D (p= 0.44) avec une moyenne de 1025 mm³ par biopsie soit 12300 mm³ par séance de biopsies en 4D (12.3 cc) contre 971 mm³ en 2D. La proportion de volume non recouvert par plusieurs biopsies (voxels biopsiés par lesquels ne passe qu'une seule biopsie) est de 56.9% en 2D contre 56.7% en 4D (p=0.98) (tableau 8).

**Tableau 8: Etude de la distribution des biopsies par la mesure du volume par biopsie et la proportion de prostate explorée par seulement une biopsie**

|  | n | Modalité | Moyenne | Ecart-type | p |
|---|---|---|---|---|---|
| **Volume moyen exploré par biopsie** | 13 | 2D | 971 mm³ | 166 | 0.44 |
|  |  | 4D | 1025 mm³ | 184 |  |
| **Proportion de prostate biopsiée par laquelle ne passe qu'une biopsie** | 13 | 2D | 0.569 | 0.1 | 0.98 |
|  |  | 4D | 0.567 | 0.09 |  |

## Discussion

La précision des biopsies sur les zones à biopsier est améliorée avec le guidage en mode 4D. Ceci illustre le fait que la représentation multidimensionnelle améliore la capacité de l'opérateur à atteindre sa cible fictive. La reproductibilité d'éxécution améliorée montre la robustesse de la technique de guidage. Il n'a par contre pas été mis en évidence d'avantage de la représentation 3D dans l'augmentation du volume exploré par la série et dans la répartition des biopsies. Ceci aurait pourtant representé un argument fort pour une augmentation de la sensibilité d'une séance de biopsies. Quelques biais ont émaillé la méthodologie de ce travail dont le recalage (fusion des volumes): une seule biopsie était faite par fantôme pour éviter que l'opérateur puisse voir ses précédentes biopsies qui auraient pu le guider. Il a donc été nécessaire d'effectuer 12 recalages par séance sur le volume de référence. Il s'en suit que le risque de réaliser un mauvais recalage est augmenté. Cependant, ces biais doivent s'annuler car sont présents quelqu'en soit le mode de guidage.

Les différentes techniques de recalage automatique développées n'ont pour l'instant pas donné de résultats.Un algorithme de recalage est en cours de développement.

Il n'est pas simple de trouver des critères de bonne qualité d'une série de biopsies prostatiques. Le but des biopsies est de détecter un foyer tumoral. La technique actuelle de biopsies échoguidées fait appel à une randomisation des biopsies au sein de la prostate. Il n'existe pas dans la majorité des cas de cible visible. Nous nous sommes créer deux objectifs: être le plus précis possible dans la réalisation d'un protocole de biopsies choisi et disperser les biopsies afin qu'elles ne se recoupent pas. La faculté à réaliser un protocole le plus précisément n'évalue donc pas la faculté des biopsies à détecter du cancer, mais la faculté d'effectuer un protocole choisi.

Un critère plus judicieux est d'évaluer la distribution des ponctions dans le volume prostatique, c'est à dire juger de la faculté à répartir les biopsies dans la prostate sans biopsier à un endroit déjà exploré. Nous avons considéré qu'une biopsie "explorait" un volume correspondant à l'intérieur d'un cylindre de 5,6mm de rayon. Ce volume ne s'est pas voulu arbitraire mais correspondant à une réalité clinique, celle de cancer de volume significatif dont la définition en vigueur est celle de Epstein (volume de 0,5 cc, score de Gleason >6). Notre technique d'évaluation de la distribution explore le recouvrement mais n'explore pas réellement l'étalement des biopsies dans le volume (figure 10).

Une séance de biopsies doit donc être jugée d'une part par la précision à mettre les ponctions où elles avaient été planifiées et d'autre part par l'aptitude à ne pas faire passer les ponctions par les mêmes points. Cette étude a permis de montrer que le premier point était amélioré par le guidage 4D mais pas le second.

La même démarche de visualisation du trajet de biopsies pourrait être applicable au vivant afin d'établir une cartographie tumorale au sein de la prostate. Le recalage des différentes biopsies permettrait d'obtenir dans un référentiel commun le trajet de toutes les biopsies.

L'intérêt clinique d'une cartographie des biopsies est évident dans 2 situations: contrôle d'une zone histologiquement douteuse vers laquelle de nouveaux prélèvement devraient être effectués ou la réalistion de biopsies itératives pour un PSA élevé ou croissant avec une première série négative. Il serait alors inutile de viser les zones préalablement biopsiés et s'orienter différemment.

Le laboratoire TIMC développe un système de guidage des biopsies vers une zone programmée dans un planning pré-procédure. Ce planning pourra être orienté sur un protocole optimisé [6] ou sur des zones préalablement biopsiées et repérées. Le but est de détecter des cancers de plus petit volume ou moins accessibles au cours d'une séance de biopsies standards. L'objectif de planning ciblé s'oppose aux protocoles de biopsies en saturation récemment proposés qui consistent à réaliser un plus grand nombre de biopsies dans la glande en essayant de couvrir le plus de volume possible.

L'étape ultime est de pouvoir réaliser un planning pré-procédure axé sur les différentes modalités d'imagerie en évaluation (spectroscopie et IRM injectée) et de réaliser en temps-réel le guidage échographique après recalage des données. Une ébauche de guidage assisté par ordinateur développé au laboratoire TIMC est en cours d'essai au CHU de Grenoble [7]

## Conclusion:

Cette étude est une étape dans l'évaluation per-opératoire du trajet des biopsies de prostate en utilisant l'échographie 3D. Elle permet de démontrer également que le guidage 3D temps-réel des ponctions sur fantôme permet une meilleure précision dans la détermination de la cible de l'aiguille et permet une meilleure reproductibilité des biopsies. Il n'a pas été mis en évidence d'argument permettant de dire que la distribution des biopsies était plus homogène au sein du volume prostatique ou que les biopsies étaient mieux réparties.

## References

1. Ravery, V., Billebaud, T., Toublanc, M., Boccon-Gibod, L., Hermieu, J. F., Moulinier, F., Blanc, E., Delmas, V., Boccon-Gibod, L.: Diagnostic value of 10 systematic TRUS-guided prostate biopsies. Eur Urol, **35:** 98-103, 1999

2. Sauer, G., Deissler, H., Strunz, K., Helms, G., Remmel, E., Koretz, K., Terinde, R., Kreienberg, R.: Ultrasound-guided large-core needle biopsies of breast lesions: analysis of 962 cases to determine the number of samples for reliable tumour classification. Br J Cancer, **92:** 231-235, 2005

3. Won, H. J., Han, J. K., Do, K. H., Lee, K. H., Kim, K. W., Kim, S. H., Yoon, C. J., Kim, Y. J., Park, C. M., Choi, B. I.: Value of four-dimensional ultrasonography in ultrasonographically guided biopsy of hepatic masses. J Ultrasound Med, **22:** 215-220, 2003

4. Polakow, J., Serwatka, W., Dobrzycki, S., JR, L. A., Janica, J., Puchalski, Z.: A new diagnostic approach to pancreatic pseudocyst fine-needle puncture: three-dimensional sonography. J Hepatobiliary Pancreat Surg, **11:** 159-163, 2004

5. Epstein, J. I., Chan, D. W., Sokoll, L. J., Walsh, P. C., Cox, J. L., Rittenhouse, H., Wolfert, R., Carter, H. B.: Nonpalpable stage T1c prostate cancer: prediction of insignificant disease using free/total prostate specific antigen levels and needle biopsy findings. J Urol, **160:** 2407-2411, 1998

6.	Chen, M. E., Troncoso, P., Johnston, D. A., Tang, K., Babaian, R. J.: Optimization of prostate biopsy strategy using computer based analysis. J Urol, **158:** 2168-2175, 1997

7.	Troccaz, J.: La chirurgie urologique assistée par ordinateur et robot. Prog Urol, **16:** 112-120, 2006